\documentclass[twocolumn,showpacs,preprintnumbers,amsmath,amssymb,prl]{revtex4-1}
\usepackage{graphicx}
\usepackage{dcolumn}
\usepackage[utf8]{inputenc}
\usepackage[ngerman, english]{babel}
\usepackage{hyperref}
\usepackage{bm}
\usepackage{xcolor}	
\usepackage{mathtools}
\usepackage{tikz}
 \usepackage{printlen}

\newcommand{\avg}[1]{\left\langle{#1}\right\rangle}

\newcommand{\C}{{\mathcal{C}}}
\newcommand{\dc}{\delta_\C}
\newcommand{\intc}[1]{\int_\C\mathrm{d}{#1}\,}

\newcommand{\chat}[1]{\ensuremath{\xop{#1}}}
\newcommand{\tempop}[3][\textstyle]{\settowidth{\dimen1}{$#1\hat{#2}$}\makebox[\dimen1][l]{$#1\hat{#2\mspace{#3}}$}}
\newcommand{\xop}[1]{{\mathchoice{\tempop[\displaystyle]{#1}{3.5mu}}{\tempop{#1}{3.5mu}}{\tempop[\scriptstyle]{#1}{3.5mu}}{\tempop[\scriptscriptstyle]{#1}{3mu}}}}
\newcommand{\cbar}[1]{\ensuremath{\xbar{#1}}}
\newcommand{\cbarbar}[1]{\ensuremath{\xbarbar{#1}}}
\newcommand{\tempbarbar}[3][\textstyle]{\settowidth{\dimen1}{$#1\bar{\bar{#2}}$}\makebox[\dimen1][l]{$#1\bar{\bar{#2\mspace{#3}}}$}}
\newcommand{\tempbar}[3][\textstyle]{\settowidth{\dimen1}{$#1\bar{#2}$}\makebox[\dimen1][l]{$#1\bar{#2\mspace{#3}}$}}
\newcommand{\xbarbar}[1]{{\mathchoice{\tempbarbar[\displaystyle]{#1}{3.5mu}}{\tempbarbar{#1}{3.5mu}}{\tempbarbar[\scriptstyle]{#1}{3.5mu}}{\tempbarbar[\scriptscriptstyle]{#1}{3mu}}}} 
\newcommand{\xbar}[1]{{\mathchoice{\tempbar[\displaystyle]{#1}{3.5mu}}{\tempbar{#1}{3.5mu}}{\tempbar[\scriptstyle]{#1}{3.5mu}}{\tempbar[\scriptscriptstyle]{#1}{3mu}}}} 
\renewcommand{\d}{{\mathrm{d}}}
\renewcommand{\i}{{\mathrm{i}}}

\newcommand{\lowerbossphantom}{\vphantom{\cbarbar{x}}}
\newcommand{\upperbossphantom}{\vphantom{\dagger}}

\newcommand{\aop}[2]{\ensuremath{\chat{c}_{#1,#2\lowerbossphantom}^{\upperbossphantom}}}
\newcommand{\cop}[2]{\ensuremath{\chat{c}_{#1,#2\lowerbossphantom}^{\dagger\upperbossphantom}}}
\newcommand{\G}[3]{\ensuremath{G_{#1#2\lowerbossphantom}^{#3\upperbossphantom}}}
\newcommand{\N}[2]{\ensuremath{N_{#1\lowerbossphantom}^{#2\upperbossphantom}}}
\newcommand{\V}[2]{\ensuremath{V_{#1\lowerbossphantom}^{#2\upperbossphantom}}}
\newcommand{\n}[2]{\ensuremath{n_{#1\lowerbossphantom}^{#2\upperbossphantom}}}
\newcommand{\E}[2]{\ensuremath{\epsilon_{#1\lowerbossphantom}^{#2\upperbossphantom}}}

\newcommand{\T}[2]{\ensuremath{T_{#1#2\lowerbossphantom}^{\upperbossphantom}}}
\renewcommand{\S}[3]{\ensuremath{\Sigma_{#1#2\lowerbossphantom}^{#3\upperbossphantom}}}
\newcommand{\h}[3]{\ensuremath{h_{#1#2\lowerbossphantom}^{#3\upperbossphantom}}}
\newcommand{\kronecker}[2]{\delta^{\upperbossphantom}_{{#1},{#2}\lowerbossphantom}}

\newcommand{\sx}{s}
\newcommand{\sxbm}{\bm{\sx}}
\newcommand{\nsx}{N_{\mathrm{s}}}

\newcommand{\spbm}{{\sxbm}^{\prime}}
\renewcommand{\sb}{\cbar{s}}
\newcommand{\sbbm}{\bm{\sb}}

\newcommand{\Sx}{\sigma}

\newcommand{\kx}{k}
\newcommand{\kxbm}{\bm{\kx}}
\newcommand{\px}{p}
\newcommand{\pxbm}{\bm{\px}}
\newcommand{\qx}{q}
\newcommand{\qxbm}{\bm{\qx}}

\newcommand{\reffig}[1]{Fig.~\ref{#1}}
\begin{document}
\bibliographystyle{apsrev}

\title{Ab initio transport results for strongly correlated fermions}

\author{\begin{otherlanguage}{ngerman}N.~Schlünzen\end{otherlanguage}}
\author{S.~Hermanns}
\author{M.~Bonitz}
\affiliation{\begin{otherlanguage}{ngerman}Institut für Theoretische Physik und Astrophysik, Christian-Albrechts-Universität zu Kiel, D-24098 Kiel, Germany\end{otherlanguage}}
\author{C.~Verdozzi}
\affiliation{Department of Mathematical Physics, Lund University, Sweden}

\pacs{05.30-d, 71.10.Fd, 67.85.-d}

\date{\today}

\begin{abstract}
Quantum transport of strongly correlated fermions is of central interest in condensed matter physics. Here, we present first-principle
nonequilibrium Green functions results using $T$-matrix selfenergies for finite Hubbard clusters of dimension $1,2,3$. We compute the expansion dynamics following a potential quench and predict its dependence on the interaction strength and particle number. We discover a universal scaling, allowing an extrapolation to infinite-size systems, which shows excellent agreement with recent cold atom diffusion experiments [Schneider {\em et al.}, Nat. Phys. {\bf 8}, 213 (2012)].
\end{abstract}
\maketitle
Particle, momentum and energy transport of strongly correlated quantum systems are of growing current interest in condensed matter~\cite{pavarini11,balzer_prb09, uimonen11}, ultracold quantum gases~\cite{schneider_np12,kajala_prl11,ronzheimer13} and dense plasmas~\cite{whitley_cpp15}.
Recently, direct measurements of quantum transport based on the expansion of ultracold atoms following a confinement quench in Hubbard-type one- and two-dimensional (1D, 2D) optical lattices with single-site resolution were reported~\cite{schneider_np12, ronzheimer13}. Also, the dynamics following a quench in lattice depth have been measured~\cite{trotzky_probing_2012, cheneau_light-cone-like_2012,will_observation_2015,zhang_observation_2012}.
On the other hand, theoretical studies of these transport processes pose fundamental challenges. 
The authors of Ref.~\cite{schneider_np12} presented numerical results for the expansion of ultracold fermions in a 2D lattice from a semi-classical Boltzmann equation model with a collision integral in relaxation time approximation (RTA) and reported good overall agreement with the experiment. At the same time they pointed out that, while the expansion can be modeled in 1D using a density matrix renormalization group (DMRG) approach~\cite{kajala_prl11, ronzheimer13}, ``[$\ldots$] so far no methods are available to calculate the dynamics quantum-mechanically in higher dimensions.''

It is the purpose of this Letter to fill this gap~\cite{CV_comment}. We present first-principle nonequilibrium Green functions simulations using the full $T$-matrix approximation applied to a fermionic Hubbard model of dimension $D=1\ldots3$. We analyze the expansion of an initially confined system and resolve the short-time dynamics of the particles demonstrating how correlations build up in finite strongly correlated inhomogeneous fermionic systems. We also analyze the long-time limit of the expansion velocity, $v^\infty_{\rm exp}$, for 1D, 2D and 3D systems, for a broad range of coupling strengths $U$ and particle numbers $N$ and observe a universal scaling $v^\infty_{\rm exp} \sim 1/\sqrt{N}$, independent of $U$ and $D$. This enables us to extrapolate our results to the macroscopic limit and directly compare with the measurements of Ref.~\cite{schneider_np12}. The agreement is excellent and achieved without any free parameters.

{\bf Nonequilibrium Green functions (NEGF) in $T$-matrix approximation (TMA).}
The NEGF are defined on the complex Keldysh contour $\C$ with time-ordering operator $T_{\C}$ as
\begin{align}
 \G{\sxbm}{\spbm}{\Sx} (z,z') = -\frac{\i}{\hbar}\avg{T_\C\aop{\sxbm}{\Sx}(z)\cop{\spbm}{\Sx}(z')}\,,
\end{align}
for lattice site indices $\sxbm=(\sx_{1},\ldots,\sx_{D}),\spbm$ and spin projection $\sigma\in\,\left\{\uparrow , \downarrow\right\}$.
The equations of motion for the NEGF are the Keldysh--Kadanoff--Baym equations (KBE) \cite{book_kadanoffbaym_qsm, bonitz-book}, 
\begin{align}
\label{eq:kbe}
&\left(\i\hbar \frac{\partial}{\partial z}\kronecker{\sxbm}{\sbbm}-\h{\sxbm}{\sbbm}{\Sx}\right)\G{\sbbm}{\spbm}{\Sx}(z,z') \\
  &\quad= \dc(z-z')\kronecker{\sxbm}{\spbm}+\intc{\cbar{z}}\S{\sxbm}{\sbbm}{\Sx}(z,\cbar{z})\G{\sbbm}{\spbm}{\Sx}(\cbar{z},z')\,,\nonumber
\end{align}
  and its adjoint (summation over $\sbbm$ is implied).
The correlation part of the selfenergy (in excess to Hartree--Fock) is computed on the $T$-matrix level and reads, for the Hubbard model \cite{puigvonfriesen09},
\begin{align}
 \S{\sxbm}{\spbm}{\textnormal{cor},\uparrow(\downarrow)}(z,z') &= \i\hbar\, \T{\sxbm}{\spbm}(z,z')\,\G{\spbm}{\sxbm}{\downarrow(\uparrow)}(z',z)\,, 
\label{eq:sigma_t}\\
 \T{\sxbm}{\spbm}(z,z') &= -\i\hbar\, U^2\, \G{\sxbm}{\spbm}{\textnormal{H}}(z,z') \label{eq:lsg} \\
 &\quad\, +\i\hbar\, U \intc{\cbar{z}} \G{\sxbm}{\sbbm}{\textnormal{H}}(z, \cbar{z})\T{\sbbm}{\spbm}(\cbar{z},z')\,, \nonumber\\
\G{\sxbm}{\spbm}{\textnormal{H}}(z,z') &= \G{\sxbm}{\spbm}{\uparrow}(z,z')\, \G{\sxbm}{\spbm}{\downarrow}(z,z')\,.\nonumber
\end{align}
Here, $T$ can be understood as an effective interaction obeying the Lippmann--Schwinger equation (\ref{eq:lsg}), e.g.~\cite{book_kadanoffbaym_qsm,kremp_ap97,semkat_jmp00}. Taking only the leading term in (\ref{eq:lsg}), which describes the interaction with a single electron pair, the TMA reduces to the second-order Born approximation (SBA). 
We underline the conserving character of this approximation \cite{book_kadanoffbaym_qsm} and, in fact, conservation of particle number and {\em total energy} is 
observed to high accuracy in all our simulations \cite{boltzmann-comment}. 

The use of this complex approximation under full nonequilibrium conditions has only recently become possible for the Hubbard model, e.g.~\cite{puigvonfriesen09,hermanns_prb14,bonitz_cpp15}. Comparisons with exact diagonalization calculations (CI) confirmed the high accuracy of this approximation \cite{puigvonfriesen10, hermanns_prb14}. 
We performed additional simulations using $T$-matrix and second-order Born selfenergies with the generalized Kadanoff--Baym ansatz (GKBA) with Hartree--Fock propagators to reduce selfconsistency effects that are known to be critical in finite systems \cite{puigvonfriesen09}.\\ 
\begin{figure}
 \includegraphics{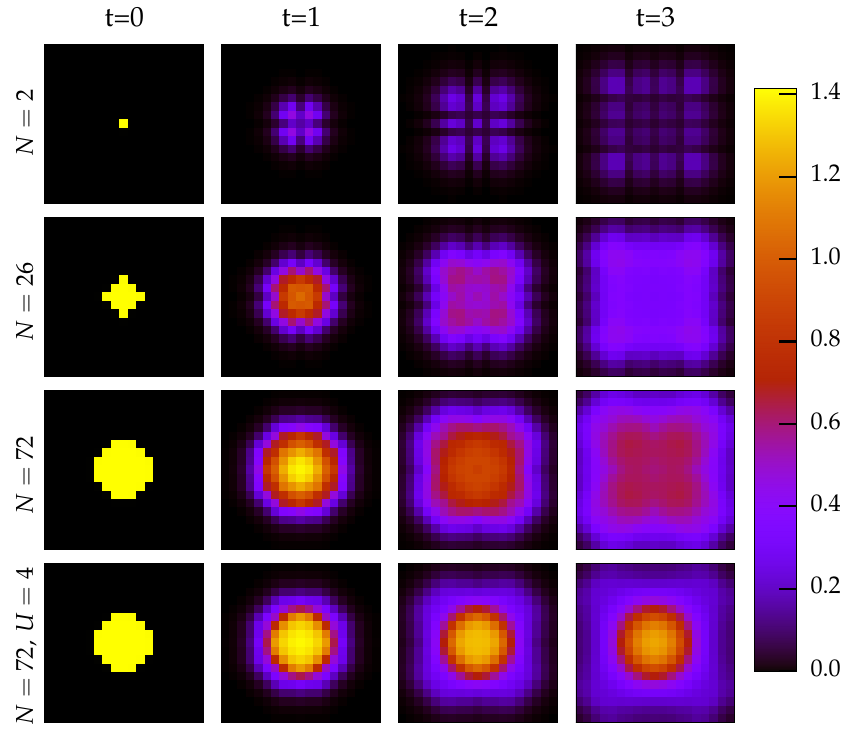}
 \caption{(Color online) Time evolution of initially circularly confined fermions in a two-dimensional $19\times 19$ Hubbard lattice for the three particle numbers $N=2, 26, 72$ for $U=1$ (upper three rows) and $U=4$ (bottom row). Results of TMA calculations. Columns correspond to four representative time steps. Color code corresponds to $\sqrt{n_{\bf s}}$.}
 \label{fig:density-overview}
\end{figure}
{\bf Numerical results.} The two-time KBE (\ref{eq:kbe}) were solved for a $D$-dimensional Hubbard model of $\N{\sxbm}{}$ sites and a step-like circular confinement $\V{}{R}(t)$ of radius $R$ (switched-off for $t>0$, as in the experiment \cite{schneider_np12}),
\begin{align}
 H(t) &= - \sum_{\avg{\sxbm,\spbm}}\sum_{\sigma=\uparrow,\downarrow}\cop{\sxbm}{\Sx}\aop{\spbm}{\Sx} + U \sum_{\sxbm}\cop{\sxbm}{\uparrow}\aop{\sxbm}{\uparrow}\cop{\sxbm}{\downarrow}\aop{\sxbm}{\downarrow} \nonumber\\
 &\quad\, + \sum_{\sxbm}\sum_{\sigma=\uparrow,\downarrow} \V{\sxbm}{R}(t) \cop{\sxbm}{\Sx}\aop{\sxbm}{\Sx} \,,
 \label{eq:hubbard}
\end{align}
 where $\avg{\sxbm,\spbm}$ denotes nearest neighbor sites. It was studied for $D=1\ldots3$ and a broad range of coupling parameters, $0 \le U \le 8$. The number of fermions, $N=N^\uparrow+N^\downarrow = 2N^\uparrow$, was varied in the range $2\dots 114$. The calculations started in the thermodynamic ground state which agrees well with the experimental conditions \cite{schneider_np12}.
After the confinement was switched off (potential quench), the diffusion of the fermion cloud was recorded.
Figure \ref{fig:density-overview} shows snapshots of a typical expansion for two couplings, $U=1,\,4$. For $U=1$, the density rapidly evolves towards square symmetry of the lattice, whereas  for $U=4$ the core region remains circular, as observed in the experiment \cite{schneider_np12}. It is apparent that, there is first a universal initial phase where diffusion is only possible for particles at the cluster edge due to Pauli blocking \cite{kajala_prl11, hermanns_prb14, lacroix14}. This is followed by a more complex evolution that strongly depends, both, on $U$ and $N$ (compare the rows in Fig.~\ref{fig:density-overview}). 

\begin{figure}
 \includegraphics{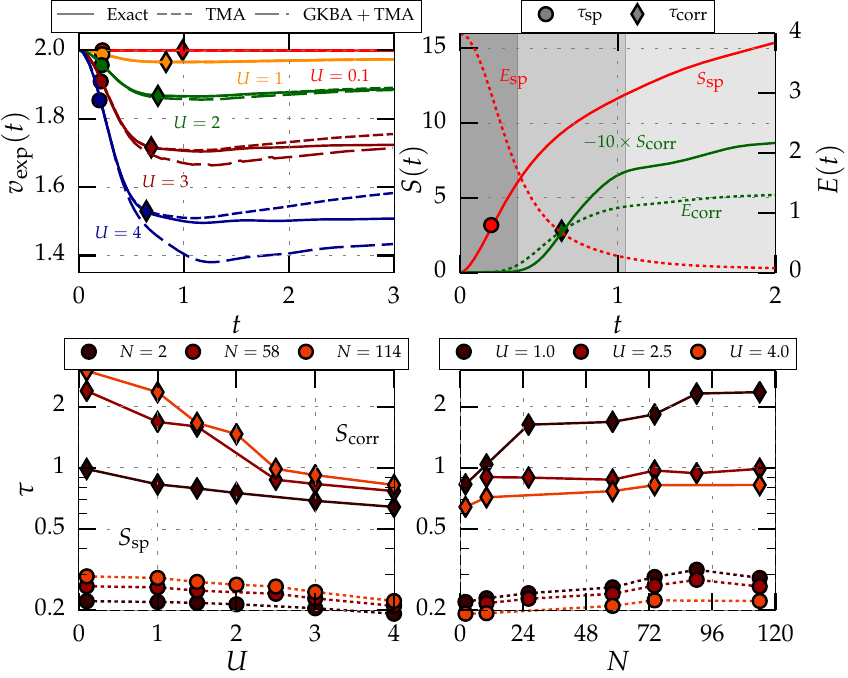}
 \caption{(Color) {\bf Top left:} Time evolution of the expansion velocity for the
$N=2$ setup of \reffig{fig:density-overview} and 
five values of $U$. The horizontal extrapolation of $v_{\rm exp}$ is used below to characterize the stationary expansion of the system. Different line styles denote the TMA, GKBA+TMA and the exact result (CI). Symbols: inflection points of the characteristic evolution quantities (see top right figure for explanation).
 {\bf Top right:} Exact evolution of the single-particle and correlation parts of the entropy and energy, for $N=2$ and $U=4$. Symbols mark the respective inflection points (positions agree with TMA results (not shown here)). Shaded areas correspond to the three phases of the evolution (see text). {\bf Bottom:} Inflection times $\tau_{\mathrm{sp}}$ (single-particle, solid lines with diamonds) and $\tau_\mathrm{corr}$ (correlation, dashed lines with circles) from TMA. {\bf right:} Dependence on $U$ for $N=2,58,114$. {\bf left:} Dependence on $N$ for $U=1.0,2.5,4.0$.}
 \label{fig:expansion}
\end{figure}

To quantify this evolution, we introduce the cloud diameter $d$, 
corrected for the initial cloud diameter, $R^2(0)$,  
\begin{align}
 d(t) &= \sqrt{R^2(t)-R^2(0)}\,, 
\label{eq:d}\\
 R^2(t) &= \frac{1}{N} \sum_{\sxbm}^{\N{\sxbm}{}} \n{\sxbm}{} (t)\, \lVert\sxbm-\sxbm_0\rVert^2 \,, 
\quad \sxbm_0 = \frac{1}{N} \sum_{\sxbm}^{\N{\sxbm}{}} \n{\sxbm}{} (0)\,\sxbm \,,
\nonumber
\end{align}
that involves the time-dependent site occupation numbers $\n{\sxbm}{}(t)$ and the static cloud center-of-mass, $\sxbm_0$. 
The upper part of \reffig{fig:expansion} shows the dynamics of the instantaneous expansion velocity, $v_{\rm exp}(t)= \frac{\d}{\d t} d(t)$ for various $U$, for the test case $N=D=2$ where exact diagonalization (CI) data are available. Our NEGF results within TMA and GKBA+TMA show good agreement with CI~\cite{acc_comment} which gives us confidence in their reliability for much larger systems and higher dimensions that are out of reach for exact methods.

The expansion velocity starts from the ideal ballistic value, $v_\mathrm{exp}(0)=\sqrt{2D}=2$, and converges to a constant smaller value, $v_\mathrm{exp}^\infty$ that monotonically decreases with $U$. The dynamics of $v_{\rm exp}(t)$ between these limits are resulting from the non-trivial interplay between single-particle and correlation effects which are straightforwardly accessible within our NEGF approach. Of particular interest are the single-particle and correlation energy ($E_{\rm sp},\,E_{\rm corr}$) \cite{hermanns_prb14} as well as 
the entanglement entropy $S=S_\mathrm{sp}+S_\mathrm{corr}$ \cite{larsson_05, friesen11},
\begin{align}
 S = \sum_{\sxbm} &-2 \left(\frac{\n{\sxbm}{}}{2} - \n{\sxbm\sxbm}{\uparrow\downarrow}\right)\mathrm{log}_2\left(\frac{\n{\sxbm}{}}{2} - \n{\sxbm\sxbm}{\uparrow\downarrow}\right) - \n{\sxbm\sxbm}{\uparrow\downarrow}\,\mathrm{log}_2\n{\sxbm\sxbm}{\uparrow\downarrow} \nonumber\\
 &-\left(1-\n{\sxbm}{}+\n{\sxbm\sxbm}{\uparrow\downarrow}\right)\mathrm{log}_2\left(1-\n{\sxbm}{}+\n{\sxbm\sxbm}{\uparrow\downarrow}\right)\,, \label{ent_ent}
\end{align}
where $\n{\sxbm\sxbm}{\uparrow\downarrow} = \avg{\cop{\sxbm}{\uparrow}\aop{\sxbm}{\uparrow}\cop{\sxbm}{\downarrow}\aop{\sxbm}{\downarrow}}$ is the double occupation of site $\sxbm$. The single-particle part, $S_\mathrm{sp}$, follows from the replacement $\n{\sxbm\sxbm}{\uparrow\downarrow} \to \n{\sxbm}{\uparrow}\n{\sxbm}{\downarrow}$, in Eq.~(\ref{ent_ent}), and the correlation part is the remainder.
The dynamics of the energy and entropy contributions allow us to identify three characteristic phases of the evolution: during the first,  $S_\mathrm{sp}$ and $E_{\rm sp}$ are built up, resulting in a decrease of $v_{\rm exp}$, see top part of~\reffig{fig:expansion}. Here, the increase of $S_\mathrm{sp}$ measures the transition from a state of independent particles ($S=0$) to an interacting many-body state. The inflection point $\tau_{\mathrm{sp}}$ (circles) of $S_\mathrm{sp}$ (and $E_\mathrm{sp}$) is representative for the time scale of this phase. Subsequently, a second phase ensues that is characterized by the saturation of $E_\mathrm{sp}$ leading to a convergence of $v_{\rm exp}$. The simultaneous build-up of correlations partly prolongs the saturation and determines the final value of $v_\mathrm{exp}$. The representative time for these processes, $\tau_{\mathrm{corr}}$, is the inflection point of $S_\mathrm{corr}$ (and $E_\mathrm{corr}$, diamonds). In the third phase, the expansion velocity and correlations are saturated, whereas the single-particle observables continue to increase with the ongoing expansion.

Both characteristic time scales show an interesting dependence on $U$, cf. upper left part of \reffig{fig:expansion}, and also on $N$.
This is further explored in the bottom part of \reffig{fig:expansion}, where we show $\tau_\mathrm{sp}$ and $\tau_\mathrm{corr}$ for varying $U$ (left) and $N$ (right). 
It is evident that $\tau_{\mathrm{corr}}$ is one order of magnitude larger than $\tau_{\mathrm{sp}}$---in striking contrast to homogeneous systems~\cite{time_scales}, and both increase with $N$. 
The reason is that entanglement entropy (and energy) are first produced at the cluster edges and the build-up continues towards the center once the outer doubly occupied sites are depopulated. For small $N$, the active regions quickly overlap impeding the production.   
On the other hand, when $U$ is increased, the effective scattering rates increase what accelerates the inward propagation of energy and entropy.

To quantify the stationary properties, we extrapolate the expansion velocity to $t \to \infty$. The result $v^\infty_{\rm exp}(U;N)$ is well suited to characterize the influence of correlations and of finite-particle number effects.
Let us now consider how the diffusion properties depend on the cloud size. To this end, we analyze the asymptotic expansion velocity, $v^\infty_{\rm exp}(U;N)$, at fixed $U$, for varying $N$. The details of the extrapolation and the calculation of the error are explained in the supplementary material~\cite{supplement}.
\begin{figure}
 \includegraphics{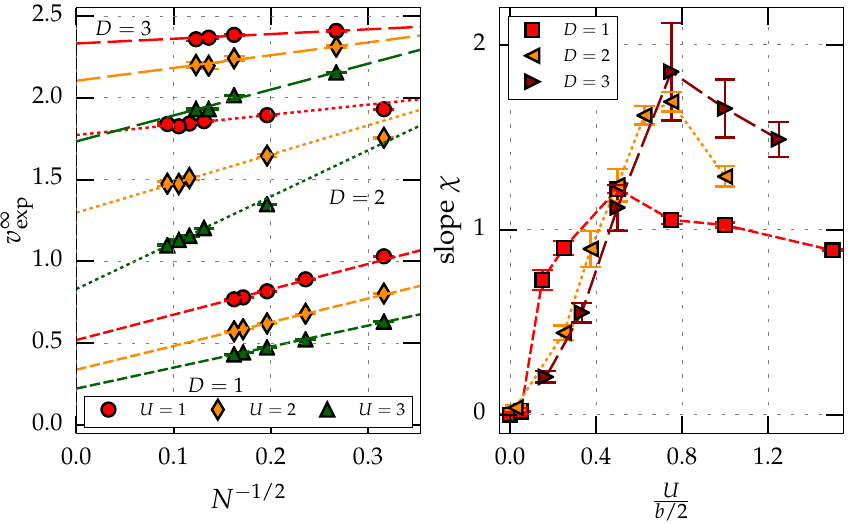}
 \caption{(Color online) \textbf{Left:} Asymptotic expansion velocity vs. particle number for $D=1\ldots3$ and $U=1\ldots3$. Symbols correspond to actual TMA data points, errors are smaller than symbol size. Dashed lines: linear extrapolation $N\to \infty$ according to Eq.~\ref{eq:extrapolation}. {\textbf{Right:}} Corresponding slope $\chi$ vs. bandwidth-normalized interaction.
}
 \label{fig:n-extrapolation}
\end{figure}
The left-hand part of \reffig{fig:n-extrapolation} shows the obtained results for $U=1\ldots3$ and different dimensions $D=1\ldots3$. Beside the above-mentioned decrease of $v^\infty_{\rm exp}$ with $U$, it also shows an increase with $D$, that is due to the enlarged number of degrees of freedom.  A striking observation is that, for all $U$ and $D$, the asymptotic expansion velocity exhibits a linear scaling with $N^{-1/2}$, 
\begin{align}
v^\infty_{\rm exp}(U;N;D) - V_{\rm exp}(U;D)  = \chi(U;D) N^{-1/2}\,,
\label{eq:extrapolation}
\end{align}
for sufficiently large $N$. The right-hand part of Fig.~\ref{fig:n-extrapolation} shows the dependency of the slope $\chi$ on the bandwidth-normalized interaction strength $U/(b/2)$ with the effective bandwidth $b=4D$. For all dimensions, $\chi$ starts at zero for vanishing $U$, increases to a maximum below $U = (b/2)$ and decreases again for further increased $U$. The $N$-independent value $\chi(0,D)=0$ is a consequence of ballistic expansion of independent particles. On the other hand, for $U \to \infty$, the doubly occupied sites are effectively frozen and the particles do not expand regardless of $N$. In-between these two limits the slope shows a qualitatively similar behavior for all dimensions,  exhibiting a steep rise, for small $U$ and a slow decrease, for large $U$.       
\begin{figure}
 \includegraphics[width=0.99\columnwidth]{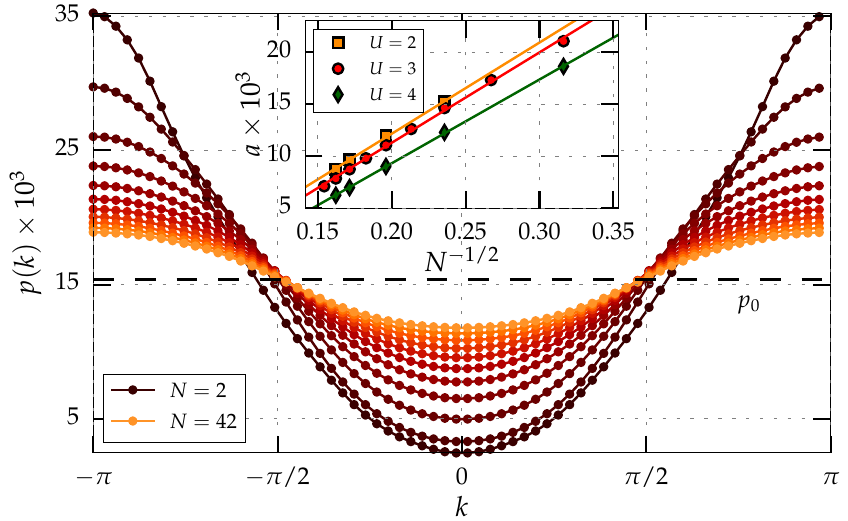}
\caption{(Color online) Momentum distribution $p(k)$ at $t=9.5$ for a 1D system of $N_\textnormal{s}=65$ for $U=3$ and different particle numbers $N=2\ldots42$ obtained with TMA. The dashed line denotes the initial distribution $p_0(k)$ which is uniform and independent of $N$. Inset shows the amplitudes $a$ of the distribution vs.  particle number for $U=2\ldots4$. Symbols correspond to actual data points, lines denote linear fits.}
 \label{fig:k_occ}
\end{figure}

To further explore the universality of the scaling with $N^{-1/2}$, we transform the Hamiltonian to momentum space~\cite{Robertbook},
\begin{align}
 \tilde H(t) &= \sum_{\kxbm}\sum_{\Sx=\uparrow,\downarrow}\E{}{}(\kxbm)\,\cop{\kxbm}{\Sx}\aop{\kxbm}{\Sx} 
 +\sum_{\kxbm,\pxbm}\sum_{\sigma=\uparrow,\downarrow}\tilde V^R_{\kxbm\pxbm}(t)\cop{\kxbm}{\Sx}\aop{\pxbm}{\Sx}\,\nonumber
\\
 &\qquad + \frac{U}{\nsx}\sum_{\kxbm,\pxbm,\qxbm}\cop{\pxbm+\qxbm}{\uparrow}\cop{\kxbm-\qxbm}{\downarrow}\aop{\pxbm}{\uparrow}\aop{\kxbm}{\downarrow},
\end{align}
where $\E{}{}(\kxbm)=-2\sum_i^D\cos (k_i)$ and $\tilde V^R_{\kxbm\pxbm}(t)$ is the transform of $V^R_{\sxbm}(t)$.
Figure \ref{fig:k_occ} shows the momentum occupation probability $p(\kxbm) = n(\kxbm)/N$ of a 1D system for $U=3$ and $2 \le N \le 42$ at the end of the simulation. For all $N$, $p(\kxbm)$ oscillates around a constant mean with an amplitude $a$, that monotonically decreases with $N$. For large $N$, we observe
\begin{align}
p({\bf k}) = p(k) = \frac{1}{\nsx} + a \cos(k)\,,
\end{align}
where the value of $a(U,N)$ is shown in the inset of \reffig{fig:k_occ} for different $U$ and we again encounter a scaling $a\sim N^{-1/2}$. The recovery 
of the same asymptotic scaling for $p(k)$ as for the expansion velocity is another indication of universal behavior. This coincidence is not surprising since $p(k)$ directly enters the kinetic energy and also determines the interaction energy.

\begin{figure}
 \includegraphics{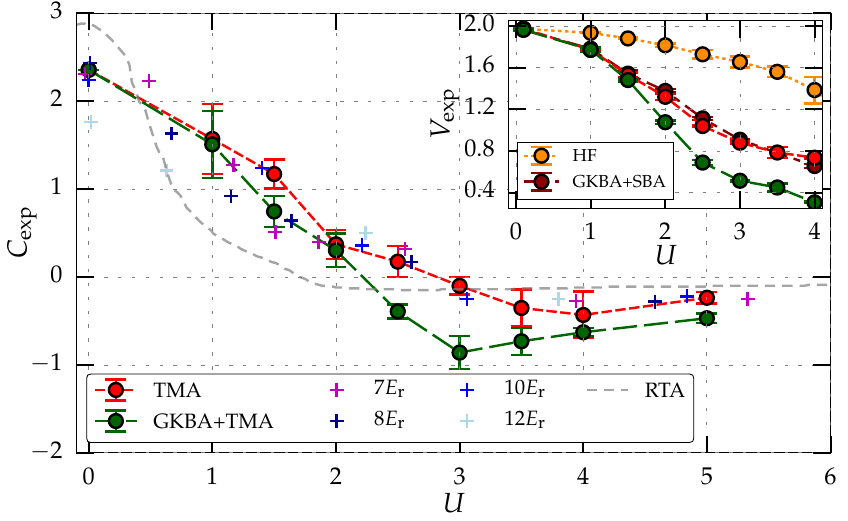}
 \caption{(Color) Extrapolated core expansion velocity $C_{\mathrm{exp}}$ from NEGF (circles with error bars), compared to the experiments (plus-signs) for different recoil energies $E_\mathrm{r}$~\cite{recoil} and the relaxation time approximation results (gray dashes) of Schneider {\em et al.} \cite{schneider_np12}. Inset: NEGF results for the asymptotic expansion velocity $V_{\mathrm{exp}}$---TMA vs. HF and SBA.}
 \label{fig:exp-comparison}
\end{figure}

The observed robust scaling makes us confident to use the extrapolation of $v^\infty_{\rm exp}(U;N)$ with respect to $N$ for quantitative predictions of the diffusion properties of strongly correlated macroscopic systems. In the inset of Fig.~\ref{fig:exp-comparison} we show the result $V_{\rm exp}(U) = \lim_{N\to \infty} v^\infty_{\rm exp}(U;N)$ as a function of $U$ and confirm the monotonic reduction of the mobility with the interaction strength. The HF approximation exhibits strong deviations which underlines the key role of correlations. Thus, having obtained macroscopic results allows us to 
compare with the experimental data that are typically obtained for large systems with $N\sim 10^5$ fermionic atoms. To this end, we compute the {\em core expansion velocity}, $C_{\rm exp}$, introduced in Ref.~\cite{schneider_np12}---the velocity of the half width at half maximum (HWHM). In the experiments, $C_{\rm exp}$ showed an interesting zero crossing, around $U=3$, after which it became slightly negative, indicating shrinkage of the central part. Even though the ``core'' may be an ambiguous definition, it is useful for quantitative comparisons. In fact, our $T$-matrix NEGF results show a surprisingly good agreement with the experimental data in the entire $U$-range, including the zero crossing at $U\approx 3$, cf. Fig.~\ref{fig:exp-comparison}. 

{\em To summarize}, we have presented first-principle NEGF results for finite ensembles of strongly correlated fermions on 1D, 2D as well as 3D Hubbard lattices. With the developed simulations within the $T$-matrix approximation, the expansion dynamics following a confinement quench could be accurately simulated. These dynamics were quantified using the expansion velocity and the relevant energy and entropy contributions, by which their coupling and particle number dependence were revealed. Our results show that full two-time quantum simulations can be successfully applied to mesoscopic fermionic systems and, via extrapolation, even can be extended to macroscopic systems. The agreement with experimental results is excellent taking into account that our theory has no free parameters. Our approach is directly applicable to other transport quantities including electrical and heat conductivity and magnetic properties. Furthermore, the fully inhomogeneous character of the simulations allows one to study the influence of the system geometry and dimension. Finally, it will be interesting to experimentally verify the observed particle number dependence of the expansion velocity and other transport properties.

We acknowledge discussions with L. Oesinghaus during the early stage of this work. This work is supported by the Deutsche Forschungsgemeinschaft via grant BO1366/9 and by grant SHP006 for super-computing time at the HLRN.


\begin{thebibliography}{26}
\bibitem{time_scales} In homogeneous many-body systems, the single-particle time scale (relaxation time) is always longer than the correlation time, e.g.~\cite{bonitz-book}. Here, the different ordering is due to the formation of the diffusion front.
\bibitem{boltzmann-comment} We note that Boltzmann-type kinetic equations in general, and the relaxation time approximation \cite{schneider_np12} in particular, conserve only the kinetic energy which makes their application to strongly correlated systems questionable, e.g.~\cite{bonitz-book}.
\bibitem{recoil} We consider the results for all values $E_\mathrm{r}$ shown in Ref.~\cite{schneider_np12} valid for comparison.
\end{thebibliography}
\end{document}


\bibliographystyle{apsrev}


\title{Supplementary Material for manuscript \\``Ab initio transport results for strongly correlated fermions''}

\author{\begin{otherlanguage}{ngerman}N.~Schlünzen\end{otherlanguage}}
\author{S.~Hermanns}
\author{M.~Bonitz}
\affiliation{\begin{otherlanguage}{ngerman}Institut für Theoretische Physik und Astrophysik, Christian-Albrechts-Universität zu Kiel, D-24098 Kiel, Germany\end{otherlanguage}}
\author{C.~Verdozzi}
\affiliation{Department of Mathematical Physics, Lund University, Sweden}

\date{\today}

\maketitle
\section{Extrapolation of the asymptotic expansion velocity}

The asymptotic expansion velocity $v_\mathrm{exp}^\infty$ is obtained from $v_\mathrm{exp}(t)$ by averaging after reaching convergence at the time $t_\mathrm{avg}$, which fulfills
\begin{align}
 \left| \frac{1}{v_\mathrm{exp}(t)} \cdot \frac{\d v_\mathrm{exp}}{\d t} (t) \right| < \epsilon\, ,
\end{align}
for all $t > t_\mathrm{avg}$ and a given small parameter ($\epsilon \ll 1$). To quantify the error of $v_\mathrm{exp}^\infty$, we use the standard deviation $\sigma\left({v_\mathrm{exp}^\infty}\right)$ of the averaging process.
The macroscopic limit $V_{\rm exp}$ is extrapolated via 
\begin{align}
v^\infty_{\rm exp}(N) =  V_{\rm exp} + \chi N^{-1/2}\, , \label{extra}
\end{align}
where $V_{\rm exp}$ and the slope $\chi$ are fit parameters. Only particle numbers $N$ larger than a cutoff value $N^\star$ are taken into account. The errors $\sigma\left({v_\mathrm{exp}^\infty}\right)$ are also included in the fit process, resulting in the final accuracy $\sigma\left({V_\mathrm{exp}}\right)$.

\section{Computation of the core expansion velocity}
The core expansion velocity $c_\mathrm{exp}^\infty$ and its macroscopic limit $C_\mathrm{exp}$ are obtained similar to~\cite{schneider_np12}. 
For $D=2$, the density distribution is averaged azimuthally for each time step. The half width at half maximum $R_\mathrm{HWHM}(t)$ of the resulting profile is used to measure the core width. Adjusted for the initial core width, $R^0_\mathrm{HWHM}$, $c_\mathrm{exp}^\infty$ is determined by fitting the resulting $R_\mathrm{HWHM}(t)$ to (cf. Eq. (6) of the main text)
\begin{align}
 R_\mathrm{HWHM}(t) = \sqrt{\left(R^0_\mathrm{HWHM}\right)^2 + \left(c_\mathrm{exp}^\infty t \right)^2} \, \label{fitcore}
\end{align}
for all $t>t_\mathrm{avg}$ with $R^0_\mathrm{HWHM}$ and $c_\mathrm{exp}^\infty$ as free fitting parameters.
Since the core of the density distribution starts to shrink for sufficiently large interaction strength $U$, we apply 
\begin{align}
 R_\mathrm{HWHM}(t) = \sqrt{\left(R^0_\mathrm{HWHM}\right)^2 - \left(c_\mathrm{exp}^\infty t \right)^2}
\end{align}
instead, following~\cite{schneider_np12}, and consider $c_\mathrm{exp}^\infty$ the speed of contraction of the core region. 

The extrapolation of $c_\mathrm{exp}^\infty$ is done as in Eq.~(\ref{extra}), resulting in the macroscopic core expansion velocity $C_\mathrm{exp}$, confirming the scaling with $1/\sqrt(N)$.  
\section{Dependence of the expansion velocity on the shape of the initial confinement}
\begin{figure}
 \includegraphics{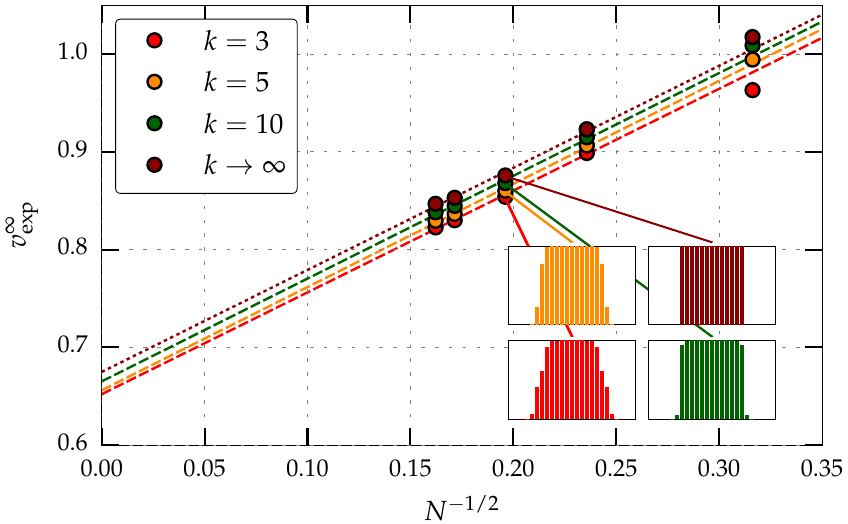}
 \caption{Dependence of the asymptotic expansion velocity on $N$ for confinement potentials of different curvature $\gamma_k(N)$. Insets show the shape of the initial density profile.
}
 \label{fig:oscillator-scaling}
\end{figure}

One may wonder whether the results of the main paper depend on the chosen steep confinement, cf. Fig. 1 of the main text.
Here, we demonstrate that the shape of the initial confinement has only a minor effect on the asymptotic expansion velocity.
To this end we use a harmonic confinement
\begin{equation}
 V(R) = \gamma_k R^2\,,
\end{equation}
with curvature $\gamma_k$. To achieve a similar shape for different $N$, we choose
\begin{align}
 \gamma_k(N) = k/N\,, 
\end{align}
for three strengths $k=3,5,10$. Together with the step-like potential ($k\to\infty$), the results are shown in Fig.~\ref{fig:oscillator-scaling}. Even though the initial density profile is affected by the curvature (see inset), the expansion velocity is not. In particular, the macroscopic limit, $V_{\mathrm{exp}}$ changes by less than $10\%$.
\section{Comparison of different many-body approximations}
\begin{figure}
 \includegraphics{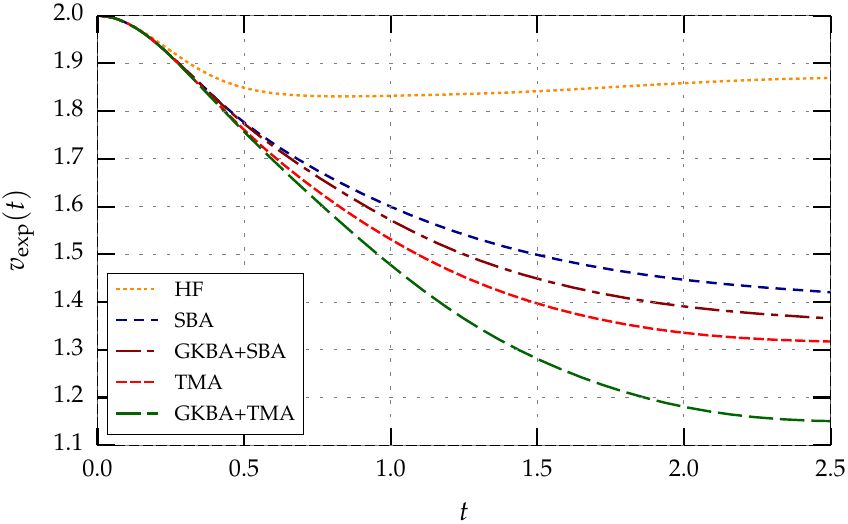}
 \caption{Time dependence of the expansion velocity for a typical example of $N=58$, $D=2$ and $U=2.5$. Comparison of selfenergy approximations: Hartree--Fock(HF), second-order Born (SBA), $T$-matrix (TMA), with and without additional generalized Kadanoff--Baym ansatz (GKBA).}
 \label{fig:methods-comp}
\end{figure}

To assess the influence of the chosen selfenergy approximation on the expansion, we computed the expansion velocity for a representative $N=58,D=2,U=2.5$ system, shown in \reffig{fig:methods-comp}. While, in Hartree--Fock approximation, $v_\mathrm{exp}$ only slightly decreases compared to the ideal velocity, the correlated methods accordantly capture a steeper decline and a later inset of convergence. Analyzing the initial behavior, one notices that, at first, the TMA and SBA start to deviate from each other. Further on, each branch divides with respect to the GKBA. In SBA, GKBA+SBA and TMA, $v_\mathrm{exp}$ converges to a value which differs by less than $10\%$. In comparison, the GKBA+TMA show a larger deviation. To classify the reliability of the approximations, we state that the TMA is superior to SBA by construction~\cite{Robertbook}. Regarding TMA and GKBA+TMA, which are both strong coupling approximations, we have shown that they differ only in the degree of selfconsistency~\cite{hermanns_prb14}. The $N=2$ test against exact results (cf. main text, Fig.~2) and the agreement with the experiment (cf. main text, Fig.~5) slightly favor TMA over GKBA+TMA, though. However, no ultimate conclusion on the preferable method can be drawn for the considered systems and quantities.     
